\documentclass{article}

\usepackage[english]{babel}
\usepackage{xcolor}
\usepackage{authblk}

\usepackage[letterpaper,top=2cm,bottom=2cm,left=3cm,right=3cm,marginparwidth=1.75cm]{geometry}

\usepackage{amsmath}
\usepackage{graphicx}
\usepackage[colorlinks=true, allcolors=blue]{hyperref}
\usepackage{amssymb}
\usepackage{subfig}

\numberwithin{equation}{section}
\numberwithin{table}{section}

\usepackage[
        backend=bibtex,
        sorting=none,
        style=phys,
        eprint=true,
        doi=false,
        biblabel=brackets
        ]{biblatex}

\bibliography{sample2}

\title{Corrections to the CRV pattern}
\author[a]{Ivano Basile}
\author[a,b]{Georgina Staudt}

\affil[a]{\small {\it {Max-Planck-Institut f\"ur Physik (Werner-Heisenberg-Institut), Boltzmannstraße 8,  85748 Garching, Germany,}}}
\affil[b]{\small{\it{Arnold-Sommerfeld-Center for Theoretical Physics, Ludwig-Maximilians-Universit\"at, 80333 M\"unchen, Germany 
    }}}

\begin{document}
\maketitle

\begin{abstract}
We investigate a pattern in the string landscape recently discovered by Castellano, Ruiz and Valenzuela, extending the analysis to subleading order in some calculable infinite-distance limits of supersymmetric compactifications. We find that in the investigated setups the proposed relation between the (gradients of the) mass gap of light towers and the species scale is satisfied. Moreover, we study an analogous relation between the species scale and the recently proposed black-hole scale, which can detect the mass gap of light species via black-hole thermodynamics. We find that, while the slope of the species scale is uniformly bounded as expected, the inner products involving the mass gap does not obey an analogous bound. Replacing the mass gap by the black-hole scale introduces subtleties, preventing us from drawing the same conclusion.
\end{abstract}

\begin{flushright}
LMU-ASC 05/25 \\
MPP-2025-32
\end{flushright}

\tableofcontents

\section{Introduction}

A recurring theme in recent research in quantum gravity has been the discovery of patterns in the string landscape, followed by a clarification of their origin rooted in physical principles. This is one of various interconnected directions spurred by the swampland program \cite{Vafa:2005ui, Brennan:2017rbf, Palti:2019pca, vanBeest:2021lhn, Grana:2021zvf, Agmon:2022thq}. This approach at the intersection of top-down and bottom-up methodologies has proven to be particularly fruitful when exploring limiting regimes of gravitational effective field theories (EFTs). In the presence of dynamical gravity, at least in $d>3$ extended spacetime dimensions, these limits are accompanied by a parametric decrease in the gravitational ultraviolet (UV) cutoff $\Lambda_\text{UV} \ll M_\text{Pl}$, where $M_\text{Pl}$ denotes the $d$-dimensional Planck scale. In string theory, the physical mechanism underlying this phenomenon is the appearance of infinite towers of species which become light \cite{Klaewer:2016kiy,Blumenhagen:2018nts,Heidenreich:2018kpg,Grimm:2018cpv,Corvilain:2018lgw,Grimm:2018ohb, Font:2019cxq, Grimm:2019wtx, Grimm:2019ixq, Gendler:2020dfp, Grimm:2020ouv, Bastian:2021eom,Klaewer:2020lfg,Cecotti:2020rjq,Klaewer:2021vkr,Lanza:2020qmt,Lanza:2020htb,Heidenreich:2021yda}. In fact, the lightest such species comprise exactly Kaluza-Klein (KK) modes pertaining to mesoscopic extra dimensions or excitations of a unique light string \cite{Lee:2018urn, Lee:2019wij, Lee:2019xtm, Klaewer:2020lfg, Basile:2022zee, Alvarez-Garcia:2023gdd, Alvarez-Garcia:2023qqj}, despite the existence of higher-dimensional membranes \cite{Alvarez-Garcia:2021pxo}, minimal volumes \cite{FierroCota:2023bsp} or non-geometric sectors \cite{Aoufia:2024awo}. This remarkable dichotomy driven by string dualities, and its extension to a general statement about quantum gravity\footnote{The logical gap between properties of the string landscape and gravitational EFTs with a consistent UV completion lies in proving ``string universality''. Evidence has been collected via swampland \cite{Kim:2019ths, Montero:2020icj, Hamada:2021bbz, Bedroya:2021fbu, Bedroya:2023tch, Bedroya:2023xue, Basile:2023blg, Bedroya:2024ubj, Herraez:2024kux, Kim:2024hxe, Kaufmann:2024gqo} and bootstrap \cite{Camanho:2014apa, Caron-Huot:2016icg, Guerrieri:2021ivu, Guerrieri:2022sod, Cheung:2022mkw, Geiser:2022exp, Cheung:2023adk, Cheung:2023uwn, Haring:2023zwu, Arkani-Hamed:2023jwn, Cheung:2024uhn, Albert:2024yap, Cheung:2024obl} methods.}, has been dubbed ``emergent string conjecture'', and refines the \emph{a priori} more general swampland distance conjecture \cite{Ooguri:2006in}, namely the notion that light towers dominate infinite-distance limits. In the spirit of the swampland program, several bottom-up motivations for the presence and nature of light towers of species at the boundary of moduli spaces (or, more generally, spaces of vacua or configurations \cite{Lust:2019zwm, Stout:2021ubb, Velazquez:2022eco, DeBiasio:2022zuh, Li:2023gtt, Basile:2023rvm, Palti:2024voy, Mohseni:2024njl}) have been recently proposed \cite{Hamada:2021yxy, Stout:2021ubb, Stout:2022phm, Calderon-Infante:2023ler, Cribiori:2023ffn, Basile:2023blg, Basile:2024dqq, Bedroya:2024ubj, Herraez:2024kux}. \\

The emergence of light towers in species limits is characterized by the mass gap $m_\text{t} \ll M_\text{Pl}$ of the lightest tower\footnote{Here we define light towers in such a way that there is no parametric gap within their mass spectra. If a finite number of species is parametrically separated from the rest of the tower, one can include them in a refined EFT without changing the story qualitatively. In some settings akin to those studied in this paper, a rigid gauge-theoretic sector does decouple \cite{Castellano:2024gwi}.}. Physics at scales $E \gtrsim m_\text{t}$ cannot be captured by a $d$-dimensional EFT, but it may be captured by a higher-dimensional EFT, whenever the dominant species are KK modes of some mesoscopic internal dimensions \cite{Castellano:2023aum, Castellano:2024bna, Bedroya:2024ubj, Aoufia:2024awo, Calderon-Infante:2025ldq}. More generally, it is conceivable that there exists some intermediate EFT description above the mass gap depending on the nature of the tower. However, in the other prominent case where the dominant species include higher-spin excitations of a light string, there is no EFT which can describe physics above $m_\text{t}$ (namely the string scale). At any rate, in all cases eventually any EFT description of gravity breaks down due to the lack of black-hole microstates, dramatic spacetime topology fluctuations and other significant manifestations of quantum gravity. Letting $\Lambda_\text{QG}$ denote the scale at which no EFT can reliably describe the physics, it follows that for KK limits $\Lambda_\text{QG}$ is the higher-dimensional Planck scale, whereas for emergent string limits $\Lambda_\text{QG}$ is the string scale. In other words, $\Lambda_\text{QG}$ is expected to weigh (field-redefinition-invariant combinations of) higher-derivative terms in the effective action which originate from genuine quantum gravity effects, rather than integrating out field-theoretic degrees of freedom. With some knowledge of the species spectrum of the theory at stake, $\Lambda_\text{QG}$ can be estimated by the upper bound $\Lambda_\text{sp}$ defined by the implicit parametric equation
\begin{equation}\label{eq:species_scale_def}
    \Lambda_\text{QG} \lesssim \Lambda_\text{sp} = \frac{M_\text{Pl}}{N(\Lambda_\text{sp})^{\frac{1}{d-2}}} \, ,
\end{equation}
where $N(E) \equiv \int_0^E \rho(\epsilon) d\epsilon$ is the (single-particle) integrated degeneracy. This bound comes from estimating the scale at which tree-level and one-loop contributions to perturbative gravitational observables become comparable \cite{Dvali:2001gx, Veneziano:2001ah, Dvali:2007hz, Dvali:2007wp, Dvali:2009ks, Dvali:2010vm, Dvali:2012uq, Basile:2024dqq, ValeixoBento:2025iqu}, and is consistent with the above considerations on $\Lambda_\text{QG}$ in the presence of light towers \cite{Castellano:2022bvr, Blumenhagen:2023yws}. The scale $\Lambda_\text{sp}$ was aptly dubbed ``species scale''. Although $\Lambda_\text{sp}$ is definitionally different from $\Lambda_\text{QG}$ \cite{Castellano:2022bvr, Blumenhagen:2023yws, Basile:2023blg, Aoufia:2024awo}, if the emergent string conjecture is true the difference between the two scales amounts to (at most) a multiplicative logarithm. Since $\Lambda_\text{QG}$ is more sharply defined, also away from species limits (where parametrically $\Lambda_\text{QG} = M_\text{Pl})$, it has become commonplace to use the informative term ``species scale'' to denote $\Lambda_\text{QG}$ directly. In this paper we shall follow this convention and, from now on, replace the symbol $\Lambda_\text{QG}$ with $\Lambda_\text{sp}$. Several detailed investigations of this scale in controlled string-theoretic settings lend support to the picture we have sketched above \cite{vandeHeisteeg:2022btw, Cribiori:2022nke, Cribiori:2023sch, Cribiori:2024qsv, vandeHeisteeg:2023dlw, vandeHeisteeg:2023ubh, Castellano:2023aum, Castellano:2024bna, Blumenhagen:2024lmo, Blumenhagen:2024ydy, Aoufia:2024awo, Calderon-Infante:2025ldq}. \\

Given this state of affairs, a strategy to make progress in understanding the physical principles underlying string theory, and its relation with quantum gravity in a broader sense, is to seek additional structures or patterns in the string landscape in order to guide further investigations. A natural place to begin is to look for relations between the physical energy scales introduced above\footnote{A complementary approach is to look for relations between the ultraviolet scales defined above and infrared quantities such as dark energy or gauge couplings \cite{Abel:2021tyt, Abel:2023hkk, Abel:2024twz, Basile:2024lcz}. This approach highlights the role of UV/IR mixing in quantum gravity \cite{Castellano:2021mmx}.}. In the context of supersymmetric string compactifications with moduli spaces, such a relation between (gradients of) the tower mass gap and the species scale in infinite-distance limits was found in \cite{Castellano:2023stg, Castellano:2023jjt, Castellano:2024bna}. An analogous result persists in the interior of vector-multiplet moduli spaces of five-dimensional supergravity as well \cite{Rudelius:2023spc}, where the species scale is replaced by the tension of a BPS string. The goal of this paper is to explore this relation in the interior of moduli spaces in four dimensions, in order to understand whether some of its structure, if any, persists beyond infinite-distance limits. In order to do so, we shall focus on settings where computations in the interior of moduli spaces are feasible due to some protection mechanism or symmetry. As we shall see, the subleading corrections to the expressions proposed in \cite{Castellano:2023stg} encode physical information on the scales at stake. In particular, they are able to distinguish between the tower mass gap and the ``black-hole scale'' introduced in \cite{Bedroya:2024uva}, which asymptotes to it. In this fashion, we are able to test the robustness of the proposal of \cite{Castellano:2023stg} under modifications to which infinite-distance limits are insensitive. \\

The paper is organized as follows. In section \ref{sec:scales_swampland} we review some background on the species scale and related aspects. In particular, in section \ref{sec:ss&top} we discuss the species scale in four-dimensional Calabi-Yau compactifications. In section \ref{sec:pattern} we present the pattern found in \cite{Castellano:2023stg, Castellano:2023jjt}, and in section \ref{sec:bh_scale} we introduce the black-hole scale defined in \cite{Bedroya:2024uva}. We begin our analysis in section \ref{sec:modular_invariance}, considering type II toroidal orbifolds where modular invariance significantly simplifies matters. We perform analogous computations in section \ref{sec:axiodilatonicsection} for the axio-dilaton modulus in the Enriques Calabi-Yau \cite{Grimm:2007tm}. In section \ref{sec:bh_scale_corrections} we discuss the role of the black-hole scale to subleading order, and highlight the differences with respect to the KK scale. We conclude in section \ref{sec:conclusions} with some final remarks.

\section{Species, towers and black holes}
\label{sec:scales_swampland}

Having introduced the notion of tower mass gap $m_\text{t}$ and species scale $\Lambda_\text{sp}$, in this section we briefly review some recent results and proposals in the context of type II compactifications on Calabi-Yau threefolds. In this setting the resulting four-dimensional EFT is $\mathcal{N}=2$ supergravity, and the internal geometry provides a powerful framework to compute certain protected higher-derivative corrections.

\subsection{The species scale and topological free energy}
\label{sec:ss&top}

In a generic (locally) Lorentz-invariant EFT, Wilson coefficients can depend on scalar fields. Therefore, if the theory has a moduli space of vacua parametrized by (some of) these scalar fields, both the tower mass gap $m_\text{t}$ and $\Lambda_\text{sp}$ can vary in moduli space. This possibility, routinely realized in string theory \cite{vandeHeisteeg:2022btw, Cribiori:2022nke, Cribiori:2023sch, vandeHeisteeg:2023dlw, vandeHeisteeg:2023ubh, Castellano:2023aum, Aoufia:2024awo, Castellano:2024bna, Calderon-Infante:2025ldq}, is particularly interesting for our purposes in this work, since it allows us to study the infinite-distance behavior of these scales in a controlled setting. Since both scales, as we have defined them in the preceding section, appear in the low-energy effective action \cite{Castellano:2023aum, Bedroya:2024ubj, Aoufia:2024awo, Castellano:2024bna, Calderon-Infante:2025ldq}, supersymmetric compactifications are particularly useful to identify them in calculable higher-derivative corrections \cite{Castellano:2023aum, vandeHeisteeg:2023dlw, Castellano:2024bna}.

More specifically, the effective 4$d$ $\mathcal{N}=2$ supergravity contains four-derivative gravitational terms of the Gauss-Bonnet type \cite{Cecotti:1992vy, Bershadsky:1993cx}
\begin{equation}\label{eq:F1_term}
    \int F_{1,\text{top}} \, \text{tr} \, R \wedge \star R \, ,
\end{equation}
which are protected against corrections by supersymmetry. The function $F_{1,\text{top}}$ of the moduli fields is the genus-one free energy of the topological string propagating in the internal Calabi-Yau threefold. It has an index-like presentation
\begin{equation}
    F_{1, \text{top}} = \frac{1}{2} \int_{\mathcal{F}} \frac{d^2\tau}{\text{Im}\tau} \, \text{tr} \, (-1)^F F_\text{L} \, F_\text{R} \, q^H \overline{q}^{\overline{H}} \, , \qquad q \equiv e^{2\pi i \tau}
\end{equation}
up to an additive moduli-independent constant\footnote{While this constant is relevant for determining the desert point \cite{Long:2021jlv, vandeHeisteeg:2022btw}, we shall neglect it in this paper, since we only care about derivatives of $F_{1,\text{top}}$. Moreover, we are interested in infinite-distance limits in which $F_{1,\text{top}}$ diverges, and thus the constant is irrelevant.}, stemming from its worldsheet origin, as well as a target-space expression in terms of Ray-Singer torsion \cite{Bershadsky:1993cx, Vafa:2024fti}
\begin{equation}
    F_{1,\text{top}} = \frac{1}{2} \sum_{p,q} (-1)^{p+q} \left(p - \frac{3}{2}\right) \left(q - \frac{3}{2}\right) \, \log \det \Delta_{\Omega^{p,q}(\text{CY}_3)} \, .
\end{equation}
In \cite{vandeHeisteeg:2022btw}, a relation between the species scale and the genus-one free energy of the topological string was proposed, tying $F_{1,\text{top}}$ directly to the number of species $N_\text{sp}$ according to
\begin{equation}
\label{specscaletopstring}
    \frac{\Lambda_{\text{sp}}}{M_{\text{Pl}}} \simeq \frac{1}{\sqrt{F_{1,\text{top}}}} \, .
\end{equation}
Since we are considering four-dimensional theories, it follows that $F_{1, \text{top}} \simeq N_\text{sp}$, at least in species limits in which both quantities are large. This identification is supported by a number of considerations and checks \cite{vandeHeisteeg:2022btw}, but one ought to keep in mind that the index-like nature of $F_{1,\text{top}}$ can bring along cancellations between bosonic and fermionic contributions. Indeed, in some limits in which the effects of a $\mathcal{N} = 4 \to \mathcal{N} = 2$ breaking are diluted, these cancellations become systematic and the above identification no longer applies. This reflects the absence of the higher-derivative term of eq. \ref{eq:F1_term}. In this sense, one expects the identification $F_{1, \text{top}} \simeq N_\text{sp}$ to be valid only when at most eight supercharges are unbroken. \\

Similar considerations can be made from a bottom-up perspective. Considering the generalized prepotential of 4$d$ $\mathcal{N}=2$ supergravity, which involves the Weyl supermultiplet, one can extract the higher-derivative terms of eq. \ref{eq:F1_term} weighed by the prepotential correction $F_{1,\text{corr}}$. As discussed in detail in \cite{Cribiori:2022nke}, the entropy of large BPS black holes receives a correction encoded by the Wald formula. This correction matches the top-down microstate counting \cite{Maldacena:1997de}, and it encodes the entropy of \emph{minimal} black holes of this type \cite{Cribiori:2022nke}. Since the tower mass gap $m_\text{t} \ll \Lambda_\text{sp}$ also generically appears in higher-derivative operators, the entropy of minimal black holes is affected by it \cite{Calderon-Infante:2025ldq, Castellano:2025ljk}. However, since the leading scaling is driven by the Planck scale which is much higher than both $m_\text{t}$ and $\Lambda_\text{sp}$, the resummed entropy can be expected to scale in the same fashion. In some examples this can be indeed shown via precise computations \cite{Calderon-Infante:2025ldq, Castellano:2025ljk}. Retracing the arguments in \cite{Dvali:2007hz, Dvali:2007wp, Dvali:2009ks, Dvali:2010vm, Dvali:2012uq}, this justifies identifying $F_{1, \text{corr}} \simeq N_\text{sp}$, leading to
\begin{equation}
\label{specscalecorrected}
    \frac{\Lambda_{\text{sp}}}{M_{\text{Pl}}} \simeq \frac{1}{\sqrt{F_{\text{1,corr}}}} \, .
\end{equation}
This relation is thus obtained via a bottom-up approach which stems from calculating higher-derivative corrections to the entropy of BPS black holes. Even though eqs. \ref{specscaletopstring} and \ref{specscalecorrected} have the same form, one ought to be careful in identifying them. On purely bottom-up grounds the respective species scales above could in principle differ, as $F_{\text{1,top}} \neq  F_{\text{1,corr}}$ in general. However, in \cite{Cribiori:2022nke} it was argued that it is not only remarkable that these two results take the same form --- it might not be a coincidence at all. Indeed, a conjecture proposed in \cite{Ooguri:2004zv} states that 
\begin{equation}
    \mathcal{Z}_{\text{BH}} = |\mathcal{Z}_{\text{top}}|^2,
\end{equation}
relating the topological partition function of a black hole to the one of the topological string. If one enforces this, the genus-one free energy of the topological string and the first correction to black-hole entropy\footnote{Higher-order effects in a similar context have been recently studied in \cite{Castellano:2025ljk}. In some cases, (minimal) BPS black holes can probe infinite-distance limits \cite{Delgado:2022dkz} and the species scale \cite{Calderon-Infante:2023uhz} even at the two-derivative level \cite{Calderon-Infante:2025pls}.} discussed in this context in \cite{Cribiori:2022nke} do yield the same species scale. For our purposes in this work, we shall identify eqs. \ref{specscaletopstring} and \ref{specscalecorrected}.

\subsection{The pattern}
\label{sec:pattern}

In this section, we will summarize the results of \cite{Castellano:2023stg, Castellano:2023jjt}, which we will examine and extend in sections \ref{sec:torussection} and \ref{sec:axiodilatonicsection}. 
As we have discussed in the preceding section, the species scale becomes small (in Planck units) as the number of species grows, which can be seen as a byproduct of the swampland distance conjecture \cite{Calderon-Infante:2023ler}. Namely, an unbounded number of species arises from infinite towers of species that become exponentially light, leading to a breakdown of the EFT. This happens when probing large distances in moduli space, and thus it is also linked to an obstruction to restoring global symmetries. Since these light towers, as well as the decay rate of their mass gap, effectively govern the behaviour and the regime of validity of the EFT, studying the nature and behaviour of these light towers is essential. \\

In this respect, the work of \cite{Castellano:2023stg, Castellano:2023jjt} highlighted a pattern arising in known string theory examples, which relates the (logarithmic) gradients of the mass gap $m_t$ of the lightest tower to that of the species scale in infinite-distance limits according to
\begin{equation}
\label{thepattern}
    \frac{\nabla m_\text{t}}{m_\text{t}} \cdot \frac{\nabla \Lambda_\text{sp}}{\Lambda_{\text{sp}}} \longrightarrow \frac{1}{d-2} \, .
\end{equation}
The gradients are taken with respect to the moduli of the EFT. Since by definition $m_\text{t} \leq \Lambda_{\text{sp}}$, one can deduce sharp bounds on how quickly towers of states become light, and correspondingly the rate at which quantum gravity effects become significant in the EFT. 
From eq. \ref{thepattern}, one can derive a lower bound on the variation rate, or slope, of the masses of the lightest tower. Denoting it by $\lambda_\text{t} \equiv \big|\frac{\nabla m_\text{t}}{m_\text{t}} \big|$,
\begin{equation}
    \lambda_{\text{t}} \geq \frac{1}{\sqrt{d-2}} \, ,
\end{equation}
which also matches the proposal of \cite{Etheredge:2022opl}. Accordingly, this leads to an upper bound for the analogous slope $\lambda_\text{sp}$ of $\Lambda_{\text{sp}}$, namely \cite{Castellano:2023jjt}
\begin{equation}
    \lambda_{\text{sp}} \leq \frac{1}{\sqrt{d-2}} \, .
\end{equation}
These bounds are consistent with the emergent string conjecture\footnote{More precisely, we are interested in asymptotically flat settings. Infinite-distance limits in anti-de Sitter exhibit more nuance \cite{Basile:2022zee, Calderon-Infante:2024oed}.} and with the considerations in \cite{vandeHeisteeg:2023ubh}, further reinforcing the link between these scales. From a worldsheet perspective, modular invariance provides a rationale for this connection \cite{Aoufia:2024awo}, but it would be interesting to further pursue this line of research beyond these corners of the landscape or provide more bottom-up motivations in the spirit of \cite{Castellano:2023stg}. \\

The pattern \ref{thepattern} is argued to hold only asymptotically. In section \ref{sec:modular_invariance} we will explore the pattern in the interior of moduli space, computing the inner product of eq. \ref{thepattern} globally and extracting the first subleading corrections in several favorable setups. This analysis will allow us to gain more insight about eq. \ref{thepattern}.

\subsection{The black-hole scale}
\label{sec:bh_scale}

Recently, \cite{Bedroya:2024uva} argued that the Gregory-Laflamme instability for black holes in compactified spacetimes, as well as the Horowitz-Polchinski instability in string theory, allows one to define a new cutoff scale for the EFT. This cutoff, dubbed ``black-hole scale'' and denoted by $\Lambda_\text{BH}$, asymptotes to the tower mass gap $m_\text{t} = m_\text{KK}$ in a decompactification limit, since the internal KK scale drives the instability. However, as we shall see, $\Lambda_\text{BH}$ can deviate from $m_\text{t}$. At any rate, the parametric inequalities
\begin{equation}\label{eq:BH_inequality}
    \Lambda_{\text{BH}} \lesssim \Lambda_{\text{sp}} \lesssim M_{\text{Pl}}
\end{equation}
hold by definition. \\

In order to see how the black-hole scale arises in settings with internal dimensions, consider a $d$-dimensional Schwarzschild black hole within $D = d+p  $ dimensions, of which $p$ are compact dimensions with length scale $R$ and $d$ are extended. In the higher-dimensional description, such a black hole lifts to a black brane wrapped around the extra dimensions uniformly. Since the Schwarzschild black hole is neutral, it cannot be charged under KK states, which would be the case if the higher-dimensional black brane were not wrapped uniformly. This implies that the black hole has some implicit knowledge of the internal space. \\

The well-known Gregory-Laflamme instability \cite{Gregory:1993vy} manifests when the mass of the black hole is low enough, at which point it undergoes a phase transition. This is because another class of solutions with lower free energy appears. These solutions are missed by the lower-dimensional EFT, and thus the lower-dimensional description of thermodynamics is incomplete. The energy scale $\Lambda_{\text{BH}}$ at which this occurs corresponds to a horizon radius $r_{\text{H}} \sim R$, but the precise relations are subject to higher-derivative corrections. Further references on spacetime instabilities are \cite{Gross:1982cv} and \cite{Reall:2001ag}. As pointed out in \cite{Bedroya:2024uva}, the Gregory-Laflamme transition scale can be defined and interpreted thermodynamically as the scale at which the entropies (or free energies in the canonical ensemble) of the higher- and lower-dimensional black holes match. When the mass decreases, the higher-dimensional black hole becomes thermodynamically favored. The scale at which this happens is again the KK scale (up to higher-derivative corrections). \\

The Gregory-Laflamme instability has a stringy counterpart in the Horowitz-Polchinski instability \cite{Horowitz:1997jc, Chen:2021dsw, Balthazar:2022hno, Ceplak:2024dxm, Bedroya:2024igb}, which involves a transition between black holes and string stars. In dimensions $d < 7$ the Horowitz-Polchinski temperature $T_\text{HP} < T_\text{H}$ is below the Hagedorn temperature at which the perturbative string description breaks down. Hence, at least in the perturbative limit, it defines a black-hole scale which once more satisfies eq. \ref{eq:BH_inequality}. Moreover, the Gregory-Laflamme and Horowitz-Polchinski transitions exhibit an interesting interplay in various dimensions \cite{Emparan:2024mbp}. \\

We stress once more that neither $T_{\text{HP}}$ nor $m_{\text{KK}}$ appear in the leading-order description of the EFT. However, they are encoded in higher-derivative corrections --- and thus in the (resummed) black-hole entropy \cite{Calderon-Infante:2025ldq, Castellano:2025ljk} --- as well as in families of two-derivative black-hole solutions in certain cases \cite{Calderon-Infante:2025pls}. It is natural to ask whether $\Lambda_{\text{BH}}$ is well-defined merely in the asymptotic boundary of moduli space or in its interior as well. Since this scale is motivated both in weak string coupling limits, as well as in large-volume decompactification limits, it seems reasonable to assume that $\Lambda_{BH}$ is well defined everywhere in moduli space. This notion could provide novel physical insight on the interior of moduli spaces, since generically one expects that the ratio $\Lambda_\text{BH}/m_\text{t}$ be moduli-dependent \cite{Bedroya:2024uva}. Nevertheless, since the ratio approaches unity at infinite distance, a similar pattern to eq. \ref{thepattern} can be considered for the black-hole scale by replacing $m_\text{t}$ with $\Lambda_\text{BH}$. In other words,
\begin{equation}
\label{patternsbh}
    \nabla \text{log} \Lambda_{\text{BH}} \cdot \nabla \text{log} \Lambda_{\text{sp}} \sim \frac{1}{d-2}
\end{equation}
asymptotically at infinite distance. Arguably, both $\Lambda_{\text{BH}}$ and $\Lambda_{\text{sp}}$ are well-defined within the moduli space. However, since $\Lambda_{\text{sp}}$ has critical points, this would imply a vanishing or a divergence in $\Lambda_{\text{BH}}$. Together with the general presence of corrections in the interior of moduli space, this motivates the inequality \cite{Bedroya:2024uva}
\begin{equation}
\label{boundsbh}
    \nabla \text{log} \Lambda_{\text{BH}} \cdot \nabla \text{log} \Lambda_{\text{sp}} \leq \frac{1}{d-2} \, ,
\end{equation}
analogous to the inequality for the (slope of the) species scale discussed in \cite{vandeHeisteeg:2022btw, vandeHeisteeg:2023dlw, vandeHeisteeg:2023ubh}. In the following, we will test this bound computing the first subleading correction to the black-hole scale and the inner product of gradients in some calculable settings. 

\section{Modular invariance and the pattern in simple examples}
\label{sec:modular_invariance}

In this section, we will test the validity of the pattern (and the associated inequality) in the interior of moduli space, as well as extract the subleading corrections in infinite-distance limits, for setups where extra dimensions are or include tori. We also discuss the black-hole scale, and its deviation from the KK scale, on one-torus decompactifications to six dimensions, thus probing the validity of eqs. \ref{patternsbh} and \ref{boundsbh}.

\subsection{The pattern on a torus}
\label{sec:torussection}

We start by examining the simple example of the isotropic torus orbifold $T^6/(\mathbb{Z}_3 \times \mathbb{Z}_3)$ discussed in \cite{Cribiori:2023sch}. Recall the K\"{a}hler potential for a single two-torus modulus $T_i$,
\begin{equation}
\label{Kähler}
    \mathcal{K} = - \text{log}[-i(T_i - \bar{T}_i)].
\end{equation}
The topological free energy can be expressed as a sum over BPS states \cite{Ferrara:1991uz}
\begin{equation}\label{eq:BPS_sum}
    F_1 = -\sum_{i=1}^{h^{11}} \sum_{(m_i,n_i)\neq (0,0)} \log \frac{|m_i + n_i T_i|^2}{-i(T_i - \bar{T}_i)} = -\sum_{i=1}^3 \text{log} [-i(T_i - \bar{T}_i) |\eta(T_i)|^4] \, ,
\end{equation}
where the index $i= 1,2,3 \equiv h^{11}$ differentiates between the three two-torus contributions. However, for now we focus on the isotropic case where $T_1 = T_2 = T_3 \equiv T$. In section \ref{sec:bh_scale_corrections} we will only consider a two-torus moduli space. Hence, we find it convenient to introduce a parameter $\gamma$ to keep track of how many moduli are varied, and eventually sent to infinity; here $\gamma=3$. Therefore, the free energy simplifies to 
\begin{equation}
\label{topfreeenergytori}
    F_1 = - \, \gamma \, \text{log}[-i(T-\bar{T})|\eta(T)|^4] \, ,
\end{equation}
and the K\"{a}hler potential similarly simplifies to
\begin{equation}
\label{Kähleriso}
    \mathcal{K} = - \gamma \, \text{log}[-i(T - \bar{T})] \, .
\end{equation}
Correspondingly, the moduli-space metric components are given by $G_{T \bar{T}} = - \, \frac{\gamma}{(T-\bar{T})^2}$. \\

Notice that from the trace over the BPS masses we omitted the contribution corresponding to a strictly massless state with $m_i = n_i = 0$. This would give an infrared divergence, which has thus been already regularized. The infinite sum is also divergent in the ultraviolet, but this can be taken care of via zeta-function regularization.
For our purposes this subtlety is relevant, since it has been argued in \cite{vandeHeisteeg:2023dlw} that the massless states of the free topological energy should not be taken into account when defining the species scale. We shall see that whether one calculates corrections to the pattern with or without these contributions changes the subleading corrections. In order to keep track of whether the massless contributions are included, we introduce a binary parameter $\beta$ such that $\beta =0$ if they are neglected, and $\beta =1$ if they are present. This amounts to using as free energy the expression 
\begin{equation}
    F_1 = - \, \gamma \, \log[(-i(T-\bar{T}))^\beta|\eta(T)|^4] \, .
\end{equation}

\subsection{Corrections for three isotropic tori}
\label{sec:isotropic_torus}

\subsubsection{Exact result}
This particular case is based on the setup of section 2.3 of \cite{Cribiori:2023sch}, which we introduced above. As advertised above, the modular invariance present in this setup allows computing the various relevant quantities globally in moduli space. The parameter $\gamma=3$ for three isotropic tori, however this factor drops out of the calculation because the inner product in eq. \ref{thepattern} contains logarithmic gradients and is therefore insensitive to constant prefactors. \\

The lightest tower in the large-$T$ limit is the BPS tower identified by $n_i=0$ in the sum of eq. \ref{eq:BPS_sum}. Hence, the mass gap in Planck units is given by 
\begin{equation}
    m_\text{t} \equiv m = \frac{1}{[-i(T-\bar{T})]^{\frac{\gamma}{2}}} \, ,
\end{equation}
which we henceforth denote by $m$ to unclutter notation. We now separate real and imaginary parts of the volume modulus according to $T=x+ iy $, and consider the large-volume limit $y \rightarrow \infty$. Hence, one obtains
\begin{equation}
    \begin{aligned}
        \partial_T \text{log} \, m = \frac {i \gamma}{4y} \, , \qquad \partial_T \text{log} \, F_1^{-1/2} &=  \frac{i \gamma}{2F_1}\left(-\frac{\beta}{2y} + \frac{\pi}{6} E_2(T)\right), \\
        \partial_{\bar{T}} \text{log}\, m = -\frac {i \gamma}{4y} \, , \qquad \partial_{\bar{T}} \text{log} \, F_1^{-1/2} &=  - \frac{i \gamma}{2F_1}\left(-\frac{\beta}{2y} + \frac{\pi }{6} E_2(\bar{T})\right).
    \end{aligned}
\end{equation}
All in all, the inner product in eq. \ref{thepattern} takes the exact form\footnote{More precisely, one ought to take the gradients and inner product in the full moduli space before restricting to the isotropic locus $T_1=T_2=T_3=T$. However, due to the diagonal structure of all quantities involved, the end result is the same.}
\begin{equation}
\label{exacttorus}
    \begin{aligned}
            (\partial_T \text{log } m )G^{T \bar{T}} (\partial_{\bar{T}} \text{log } F_1^{-1/2}) &+(\partial_{\bar{T}} \text{log } m )G^{T \bar{T}} (\partial_{T} \text{log } F_1^{-1/2}) \\ = \frac 12 \frac{1}{\text{log} [(2y)^\beta |\eta(T)|^4]} \bigg(\beta &- \frac{\pi y}{6} (E_2(T) + E_2(\bar{T})) \bigg) \, .
    \end{aligned}
\end{equation}
In the previous calculations, we have used the Dedekind eta function $\eta(T)$ and the Eisenstein series $E_2(T)$. As both of them will be crucial in the calculation of the large-$T$ limit and the subleading corrections, a brief review of their properties and their large-argument behavior is provided in appendix \ref{dedekind}. For a more comprehensive review also in the context of the topological free energy, see \cite{Grimm:2007tm}. 

\subsubsection{The asymptotics}

Asymptotically as $y \to \infty$, and including the next-to-leading order behavior, eq. \ref{exacttorus} reduces to
\begin{equation}
\label{asypmt1}
    \begin{aligned}
            \frac 12 & \left(1 - {\color{red}\beta \frac {3}{\pi y} (1 - \text{log}(2y) ) } - 12 e^{-2 \pi y}(e^{2 \pi i x} + e^{- 2 \pi i x})  - \frac{6}{\pi y} e^{-2 \pi y}(e^{2 \pi i x} + e^{- 2 \pi i x}) \right) \, .
    \end{aligned}
\end{equation}
In the above expression we have colored in red the terms proportional to $\beta$ for the reader's convenience, as we shall do in the remainder of the paper. We see that in the limit $y \rightarrow \infty$ one recovers the pattern proposed by \cite{Castellano:2023jjt, Castellano:2023stg}. However, the direction from which the limit $\frac{1}{2}$ is approached depends on whether one includes the massless contributions in the computation. If $\beta = 0$, the red term in eq. \ref{asypmt1} vanishes, and the subleading term is $- 12 e^{-2 \pi y}(e^{2 \pi i x} + e^{- 2 \pi i x})$. On the other hand, if one also accounts for the massless states, the subleading correction is $\frac {3}{\pi y}\text{log}(2y)$, which is not only parametrically larger, but is also positive, whereas the sign of the other corrections depends on the value of $x$. In particular, for $x=0$ the limit is approached from below only if $\beta = 0$. If $x=\frac{1}{2}$, the limit is approached from above regardless. Here we see that, although the remaining corrections are exponentially suppressed, they play a qualitatively important role in this analysis. This finding resonates with the detailed analysis of Seiberg-Witten limits in \cite{Castellano:2024bna}, where exponentially suppressed corrections are similarly crucial for the metric and inner product to be well-defined\footnote{We are grateful to A. Castellano for pointing this out to us.}.

\begin{figure}[htp!]
    \centering
    \subfloat[Inner product of logarithmic gradients along the line $x=0$. Massless contributions are excluded.]{%
        \includegraphics[width=0.45\textwidth]{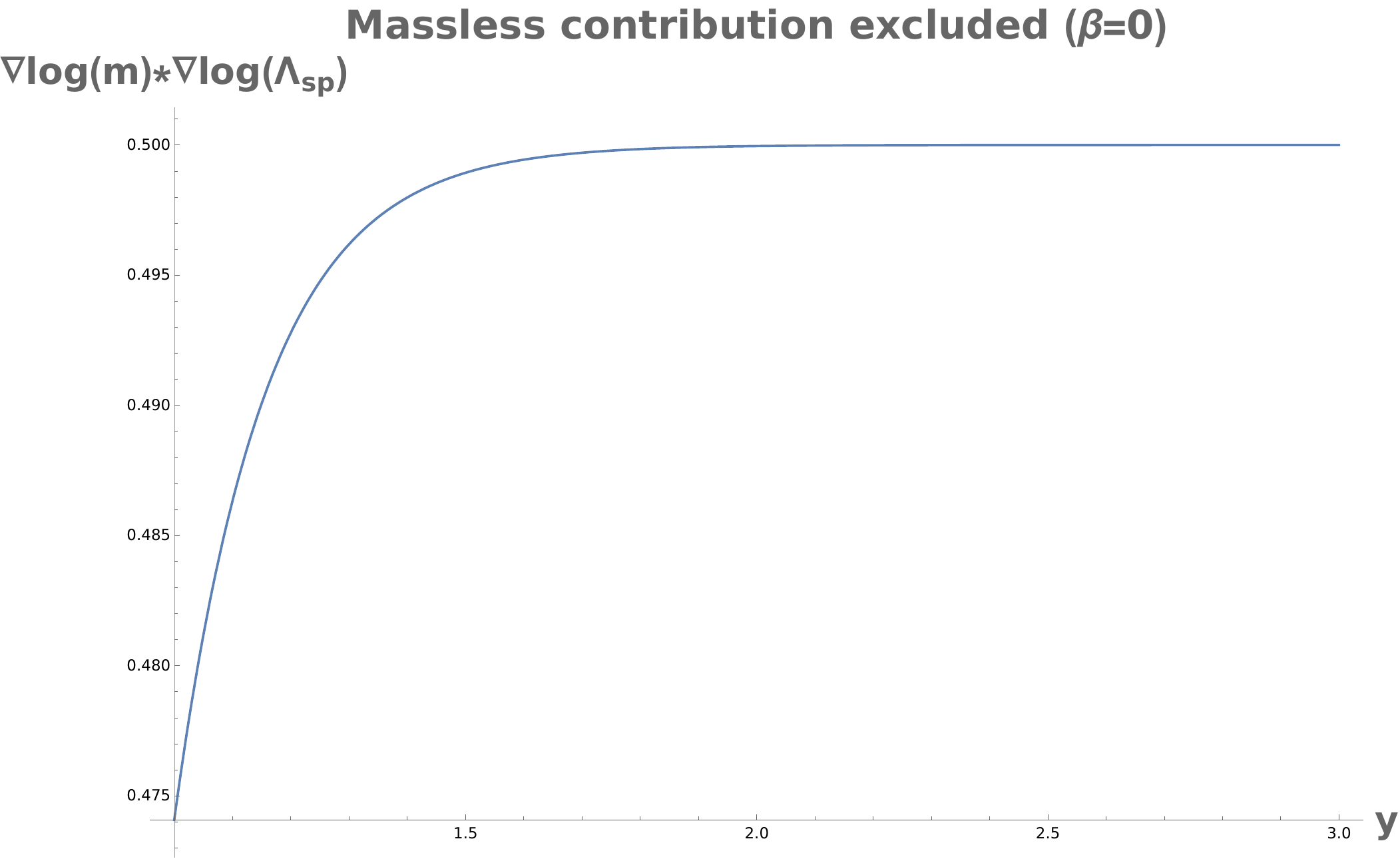}%
        \label{fig:a_2d}%
        }%
    \hspace{25pt}%
    \subfloat[Inner product of logarithmic gradients along the line $x=0$. Massless contributions are included.]{%
        \includegraphics[width=0.45\textwidth]{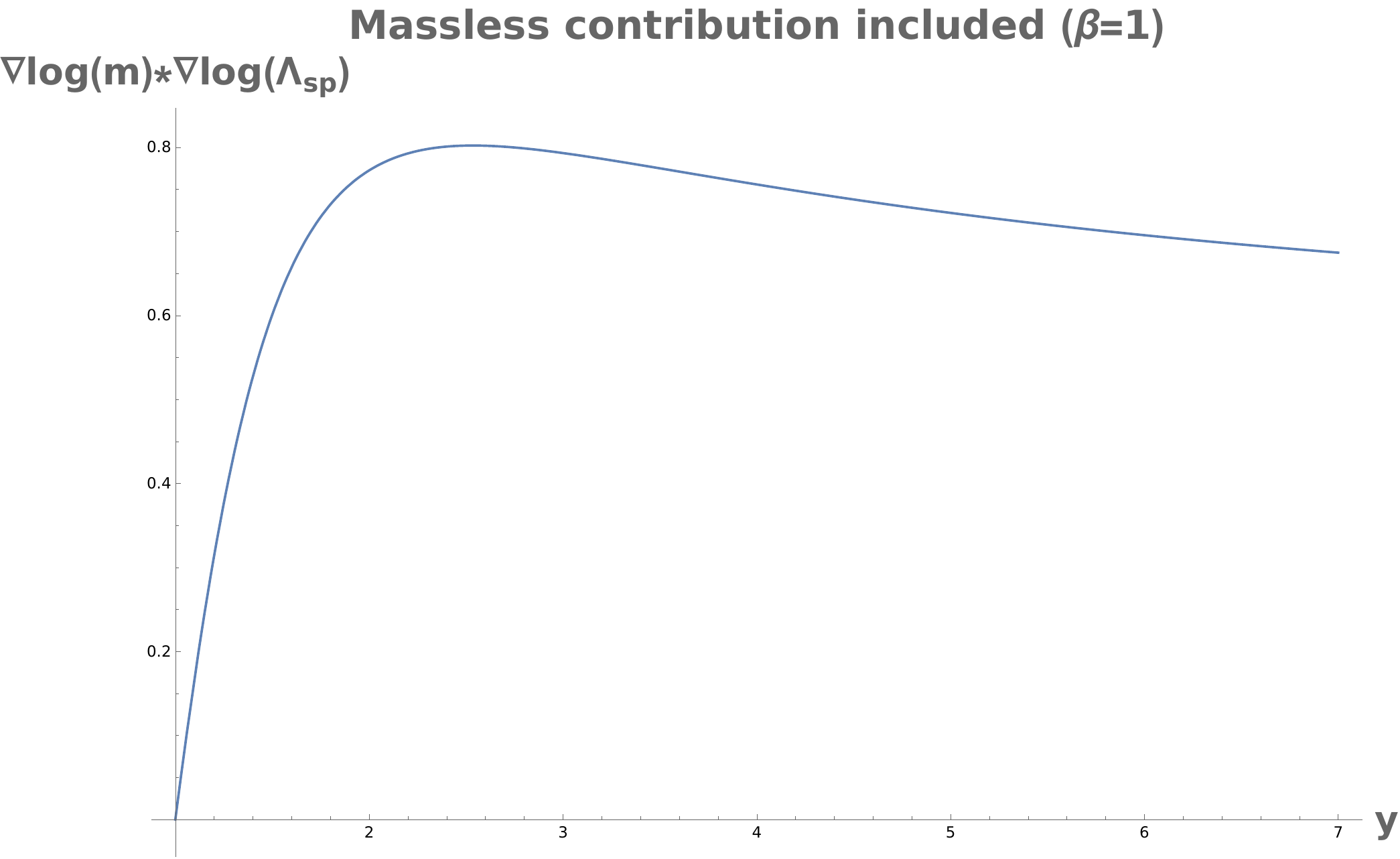}%
        \label{fig:b_2d}%
        }%
    \caption{}
\end{figure}

\begin{figure}[htp!]
    \centering
    \subfloat[Inner product of logarithmic gradients in the fundamental domain. Massless contributions are excluded.]{%
        \includegraphics[width=0.45\textwidth]{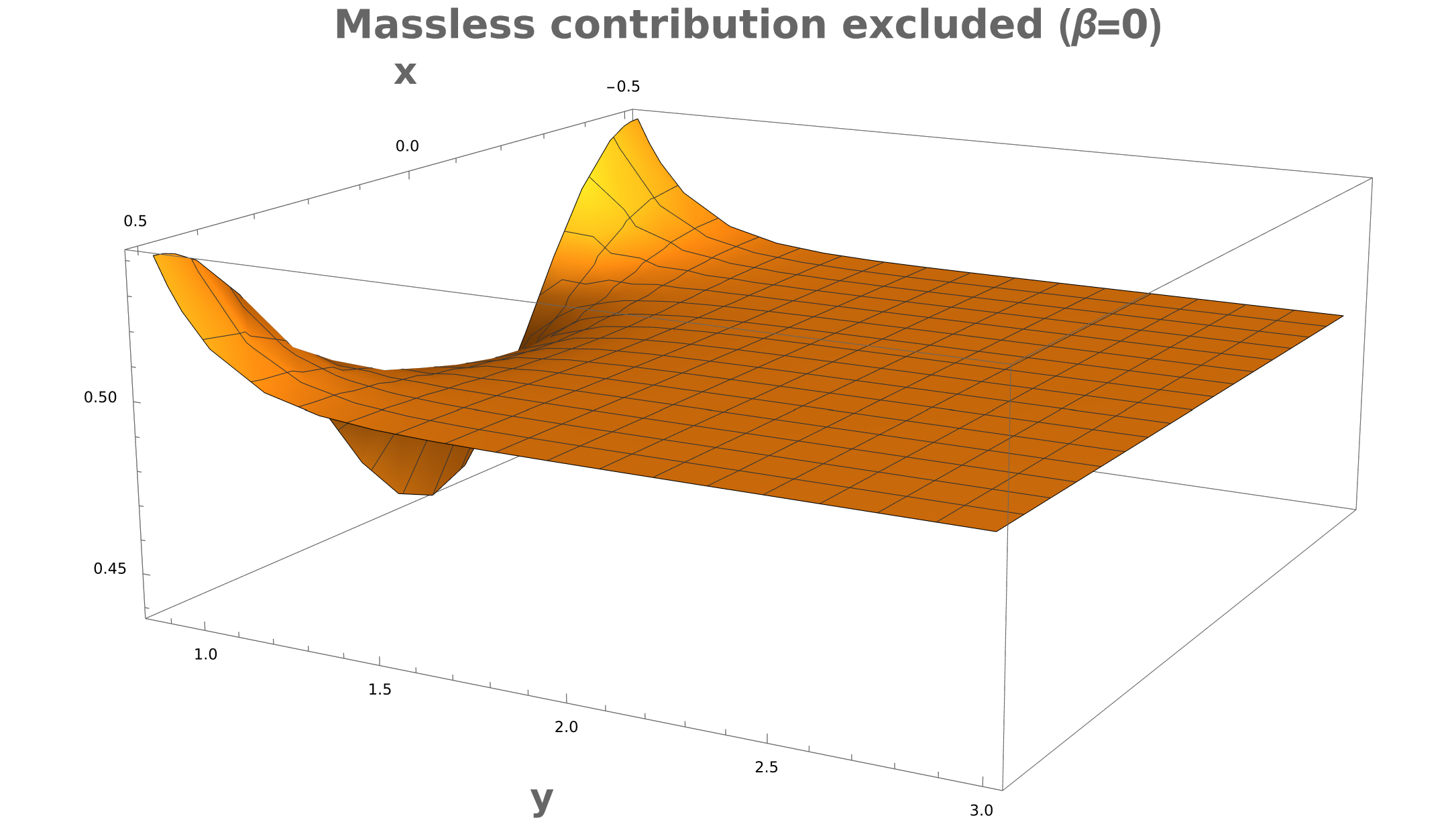}%
        \label{fig:a_3d}%
        }%
    \hspace{25pt}%
    \subfloat[Inner product of logarithmic gradients in the fundamental domain. Massless contributions are included.]{%
        \includegraphics[width=0.45\textwidth]{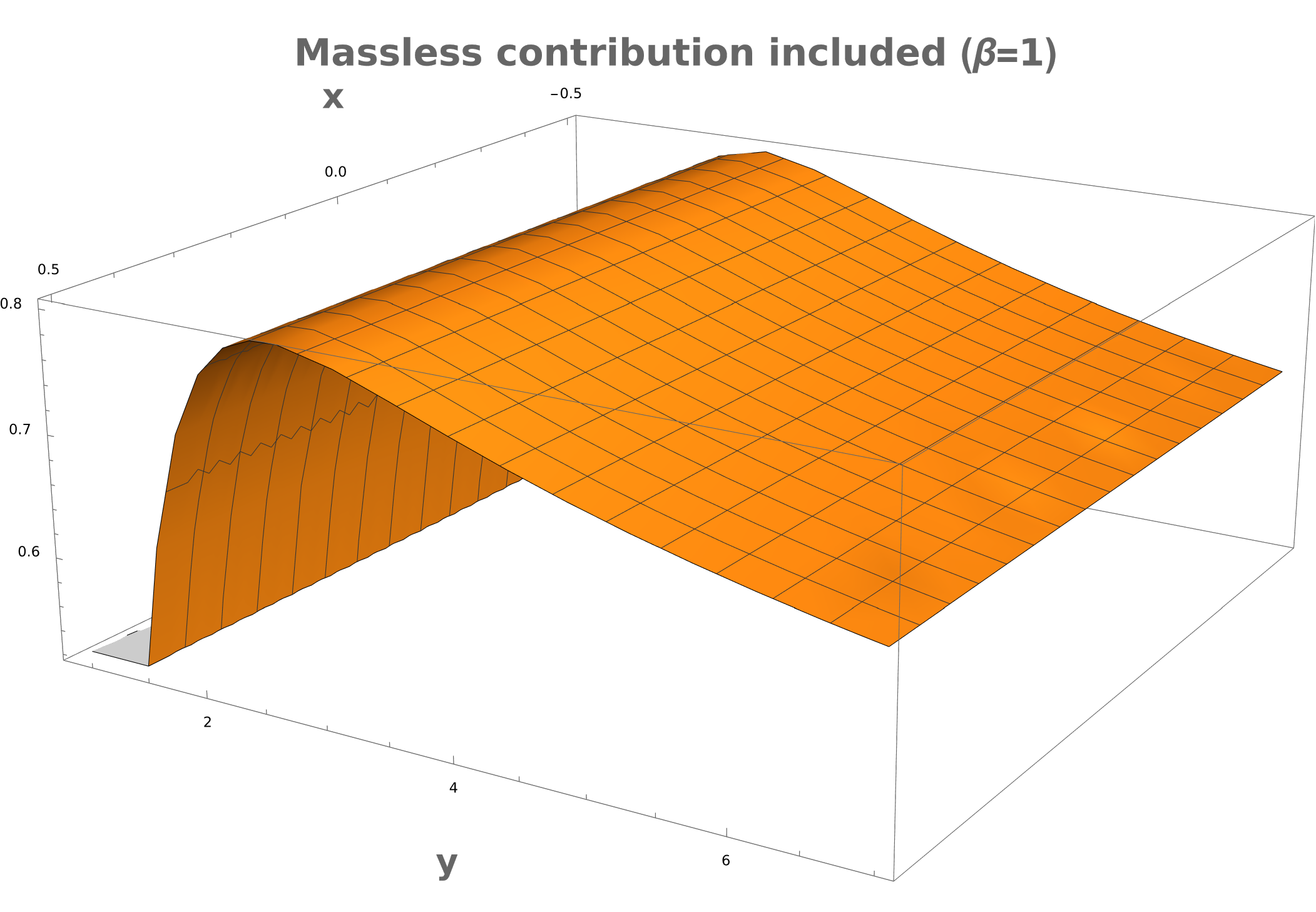}%
        \label{fig:b_3d}%
        }%
    \caption{}
\end{figure}

The above results imply that, in this limit, the inequality of eq. \ref{boundsbh} cannot be valid if the black-hole scale coincides with, or is replaced, by the tower mass gap. This asymptotic analysis can be visualized in the exact result depicted in figures \ref{fig:a_2d} and \ref{fig:b_2d} for $x=0$, and figures \ref{fig:a_3d} and \ref{fig:b_3d} for the full fundamental domain.

However, as pointed out in \cite{vandeHeisteeg:2023dlw} in the context of emergent string limits in six-dimensional F-theory, the isotropic decompactification limit with $\gamma = 3$ may restore too much supersymmetry for the identification in eq. \ref{specscaletopstring} to be reliable, as discussed in \cite{vandeHeisteeg:2023dlw}. More precisely, the contributions to $F_1$ may arise from such a sector. Indeed, computing the square $\frac{|\nabla\Lambda_\text{sp}|^2}{\Lambda_\text{sp}^2}$ with this identification, and $\beta = 0$ as in \cite{vandeHeisteeg:2023dlw}, one finds a very similar plot to figure \ref{fig:a_3d}, except that the asymptotic value is $\frac{1}{2\gamma}$. In other words, $\gamma > 1$ is inconsistent with the asymptotic values dictated by the emergent string conjecture and with the bound on the slope discussed in \cite{vandeHeisteeg:2023dlw}, indicating that in these limits $F_1$ features a larger extent of systematic cancellations. \\

\begin{figure}[htp!]
    \centering
    \subfloat[Slope of the species scale for $x=\frac{1}{2}$ and $\gamma=1$. Massless contributions are excluded.]{%
        \includegraphics[width=0.45\textwidth]{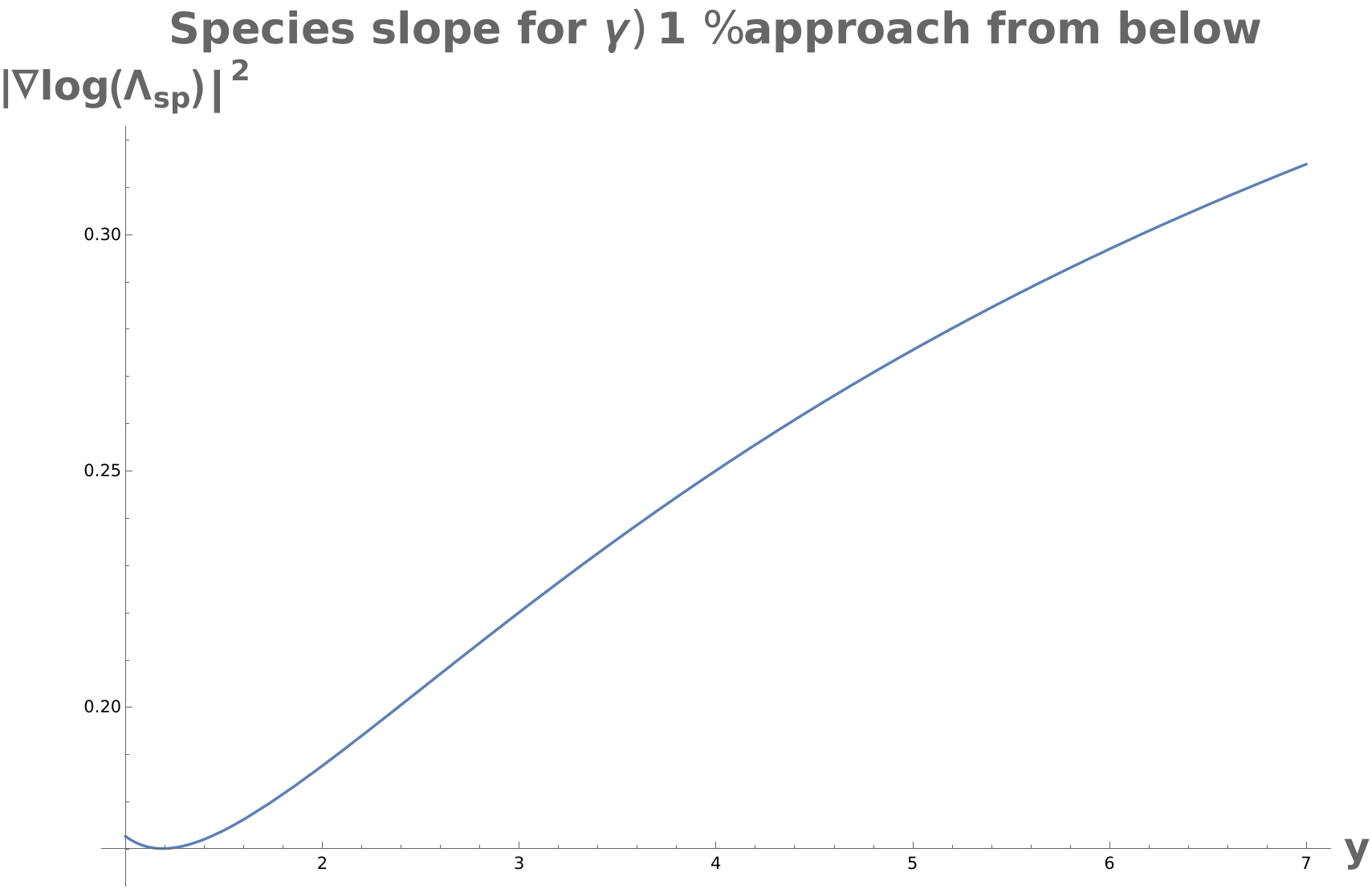}%
        \label{fig:species_slope_2d}%
        }%
    \hspace{25pt}%
    \subfloat[Slope of the species scale for a $\gamma=1$ limit in the fundamental domain. Massless contributions are excluded.]{%
        \includegraphics[width=0.45\textwidth]{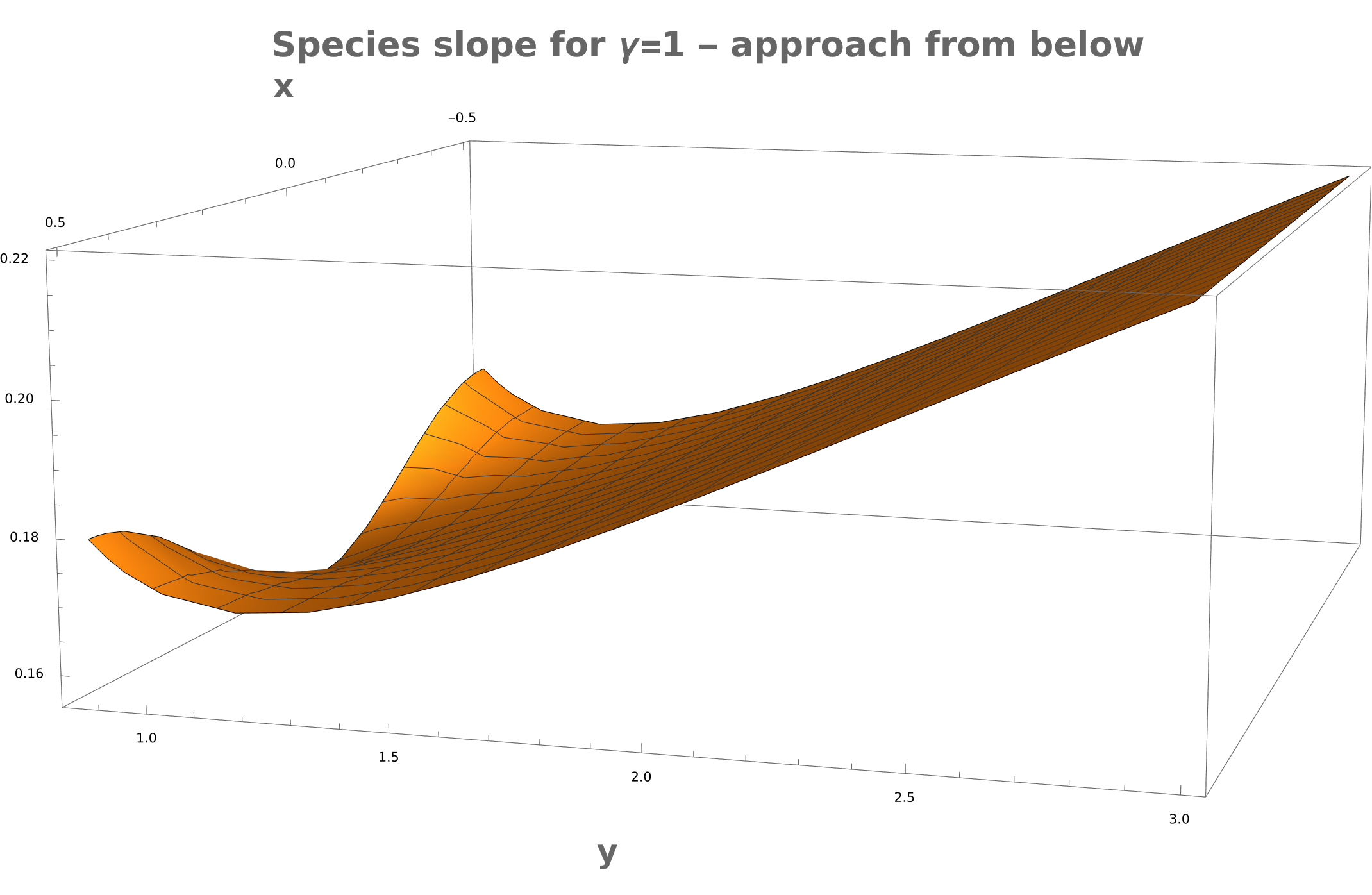}%
        \label{fig:species_slope_3d}%
        }%
    \caption{}
\end{figure}

The preceding considerations thus lead us to consider limits with $\gamma = 1$. In general, we can write
\begin{equation}
    \Bigg|\frac{\nabla \Lambda_\text{sp}}{\Lambda_\text{sp}}\Bigg|^2 = \frac{1}{2} \, \frac{\sum_{i=1}^3 |\frac{\pi y_i}{3}E_2(T_i)|^2}{(\sum_{i=1}^3 \log |\eta(T_i)|^4)^2} \longrightarrow \frac{1}{2\gamma} \, ,
\end{equation}
and for limits with $\gamma = 1$, where only, say, $y_1$ is large, the exact result now does approach the correct value $\frac{1}{2}$ from below, as shown in figures \ref{fig:species_slope_2d} and \ref{fig:species_slope_3d}.

The inner product of eq. \ref{thepattern} in this case takes the very similar form
\begin{equation}\label{eq:pattern_general_gamma}
    \frac{\nabla m}{m} \cdot \frac{\nabla \Lambda_\text{sp}}{\Lambda_\text{sp}} = -\frac{1}{2} \, \frac{\sum_{i=1}^3 \frac{\pi y_i}{3} \text{Re} \, E_2(T_i)}{\sum_{i=1}^3 \log |\eta(T_i)|^4} \longrightarrow \frac{1}{2} \, .
\end{equation}
However, even for $\gamma = 1$ limits, the result can approach the limiting value from above or below. For instance, for $T_2 = T_3 = \frac{1}{2} + 2i$, the resulting plot looks very similar to figure \ref{fig:a_3d}. Therefore, for $\gamma = 1$ limits the bound of eq. \ref{boundsbh} does not seem to hold with $m$ in place of the black-hole scale, although the slope of the species scale is bounded as in \cite{vandeHeisteeg:2023dlw}. In section \ref{sec:bh_scale_corrections} we shall investigate this bound with $m$ replaced by the black-hole scale. As we shall see, the interplay of corrections will be more subtle, preventing us from drawing the above conclusions.

\subsection{The axio-dilaton case}
\label{sec:axiodilatonicsection}

As a second example, in this section we shall consider this case, studying the pattern of eq. \ref{thepattern} in type II string theory compactified on the Enriques Calabi-Yau. This space is the quotient $(\mathbb{T}^2\times \mathrm{K3})/\mathbb{Z}_2$, where the free $\mathbb{Z}_2$ involution inverts the coordinates on the torus. Since its holonomy group is $SU(2)\times \mathbb{Z}_2 \subset SU(3)$, this compactification yields a four-dimensional theory with $\mathcal{N}=2$ supersymmetry. See \cite{Grimm:2007tm} for more details on the Enriques Calabi-Yau in string compactifications. 
In this setting, we consider the masses of BPS particles obtained wrapping D0, D2 and D4-branes (and D6-branes as the magnetic duals of D0-branes) around minimal cycles. We are interested in the limit in which the volume of the torus is large. Therefore, in four-dimensional Planck units, the BPS particles with the smallest mass (gap) depend on this modulus only via the inverse volume factor due to the conversion from string units. Letting $T^a$ and $S$ denote the $K3$ and torus moduli respectively, and $C^{ab}$ the intersection form on the Enriques $K3$ fiber \cite{Grimm:2007tm}, the relevant masses are thus given by
\begin{equation}
\label{ECYmasses}
    m_{\text{BPS}} = e^{-K_{cl}/2} |\mathcal{Z}_{\text{IIA}}| \propto \frac{1}{\sqrt{\mathcal{V}(X_3)}} \, , \qquad \mathcal{V}(X_3) = C_{ab} \, \text{Re}(T^a) \text{Re}(T^b) \, \text{Re}(S) \, .
\end{equation}
Introducing the parameter $\beta$ as in the preceding section, the topological free energy is given by 
\begin{equation}
    F_1=F^{(1)}(T, \bar{T}) - 6 \,\text{log}((S+\bar{S})^\beta |\eta(S)|^4) \, ,
\end{equation}
where the $K3$ contribution is
\begin{equation}
\label{F_T}
    F^{(1)}(T, \bar{T}) = -2 \, \log\left(\frac 12 C_{ab}(T^{a}+ \bar{T}^{a})(T^{b}+ \bar{T}^{b})\right) - \, \log|\Phi(T)|
\end{equation}
with $\Phi$ the Borcherds product \cite{Grimm:2007tm}. We shall now take the modulus $S = x + i y$, specifically its real part $x$, to be large. Thus eq. \ref{F_T} will not matter when we derive the expressions. Finally, the K\"{a}hler potential is given by
\begin{equation}
    K = - \log \left(\frac 12 (T^{a}+ \bar{T}^{a})(T^{b}+ \bar{T}^{b}) (S+\bar{S})\right) .
\end{equation}

\subsection{Corrections to the axiodilatonic case}
\label{sec:axiodilaton_corrections}

\subsubsection{Exact result}

Varying the axio-dilaton $S$, all that matters for the calculation of the inner product of logarithmic gradients is the $S$-dependent prefactor in eq. \ref{ECYmasses}. Considering D0-branes for concreteness,
\begin{equation}
    m \propto \frac{1}{\sqrt{(S+\bar{S})}} \, ,
\end{equation}
so that the calculation is essentially identical to the one in the preceding section with the parameter $\gamma=1$. For the calculation we follow the conventions in \cite{Grimm:2007tm} for the modular functions of the axio-dilaton, as reviewed in the appendix. One obtains
\begin{equation}
    \begin{aligned}
        \partial_S \text{log } m = -  \frac{1}{4x} \, , \qquad \partial_S \text{log } F_1^{-1/2} &= - \frac 12 \frac{1}{\text{log}[(2x)^\beta |\eta^2(S)|^2]}\left(\frac{\beta}{2x} - \frac{1}{12} E_2(S)\right), \\
        \partial_{\bar{S}} \text{log } m = -  \frac{1}{4x} \, , \qquad \partial_{\bar{S}} \text{log } F_1^{-1/2} &= - \frac 12 \frac{1}{\text{log}[(2x)^\beta |\eta^2(S)|^2]}\left(\frac{\beta}{2x} - \frac{1}{12} E_2(\bar{S})\right) .
    \end{aligned}
\end{equation}
Therefore, the inner product in the axio-dilaton direction is 
\begin{equation}
\label{exactaxiodilaton}
    \begin{aligned}
            (\partial_S \text{log } m )G^{S \bar{S}} (\partial_{\bar{S}} \text{log } F_1^{-1/2}) &+(\partial_{\bar{S}} \text{log } m )G^{S \bar{S}} (\partial_{S} \text{log } F_1^{-1/2}) \\ = \frac 14 \frac{1}{\text{log} [(2x)^\beta |\eta(S)|^4]} \bigg(2\beta &- \frac{x}{6} (E_2(S) + E_2(\bar{S})) \bigg).
    \end{aligned}
\end{equation}
This result is not quite exact, since the $K3$ moduli have to be included in the gradient. The leading asymptotics is unaffected, but the subleading correction is.

\subsubsection{The asymptotics}

If one takes $x$ to be large, eq. \ref{exactaxiodilaton} can be approximated by
\begin{equation}\label{eq:asymp_S}
    \frac 12 \left(1- {\color{red} \beta \frac{6}{x} (1- \text{log}(2x)) }- 12 e^{-x}(e^{- i y} + e^{i y}) - \frac{12}{x} e^{-x}(e^{- i y} + e^{i y}) \right) .
\end{equation}
Expectedly, this is the same asymptotic relation as in eq. \ref{asypmt1} up to some numerical factors. Therefore, the implications are very similar: whether the limit is always approached from above or, depending on $y$, from below, significantly depends on whether one includes the massless states. If the massless states are included, they always provide the leading correction, and the limit is approached from above. \\

However, as shown in \cite{vandeHeisteeg:2023dlw} and in the preceding discussion, one cannot neglect the $K3$ moduli in the subleading corrections. To see what this contribution looks like relative to eq. \ref{eq:asymp_S}, we use that $F_1$ and $\log m$ are additive, and the metric is block-diagonal. Hence, the additional contribution to the full inner product has the schematic form
\begin{equation}\label{eq:K3_corrections}
    \partial_T \log m(T) \, G^{T \bar{T}} \, \frac{\partial_{\bar{T}} F^{(1)}(T,\bar{T})}{F_1} \sim \frac{\mathcal{O}(1)}{x}
\end{equation}
as $x \to \infty$. Therefore, while $\beta=1$ once again provides the leading correction, for $\beta=0$ we would need to study the $K3$ sector in more detail. We leave further investigation to future work, as well as more general settings of this type where no exact results are available. For the time being, we emphasize that, for the absolute square $\frac{|\nabla \Lambda_\text{sp}|^2}{\Lambda_\text{sp}^2}$, the gradient with respect to the torus modulus is a lower bound for the full quantity. As shown in \cite{vandeHeisteeg:2023dlw}, this expression respects the upper bound of $\frac{1}{2}$.

\section{Corrections to the black-hole scale}
\label{sec:bh_scale_corrections}

Our results on the CRV pattern in the interior of moduli space indicate that if an inequality of the form of eq. \ref{boundsbh} holds, it would indeed require replacing the tower mass scale $m$ with the black-hole scale, as proposed in \cite{Bedroya:2024uva}. In this section we investigate this proposal. In order to derive a meaningful black-hole scale $\Lambda_\text{BH}$, we follow \cite{Bedroya:2024uva} and include subleading corrections relative to the KK scale. As we have discussed above, in the decompactification limits we are interested in, this scale is defined by the Gregory-Laflamme transition. Since the topological free energy is linked to higher-derivative corrections to the entropy, it is imperative that we include the latter in order to detect a deviation between $\Lambda_\text{BH}$ and $m$. We will therefore compare the entropies of four-dimensional and higher-dimensional black holes including the Gauss-Bonnet correction as in \cite{Clunan:2004tb}. Since this definition involves unprotected quantities, our results are inevitably only reliable in the relevant infinite-distance limits. \\

Since the logarithmic derivatives are insensitive to constant prefactors, we can write the leading-order entropies of a $d$-dimensional and $D$-dimensional black hole of mass $M$ as
\begin{equation}
    S^{(0)}_d = c_d \left(\frac{M}{M_\text{Pl,d}}\right)^{\frac{d-2}{d-3}} \, , \quad S^{(0)}_D = c_D \left(\frac{M}{M_\text{Pl,D}}\right)^{\frac{D-2}{D-3}} = c_D \left(\frac{M}{M_\text{Pl,d}^{\frac{d-2}{D-2}}}\right) \text{Vol}_p^{\frac{1}{D-3}} \, ,
\end{equation}
where for isotropic toroidal compactifications from $D=d+p$ to $d$ dimensions the internal volume is $\text{Vol}_p = (2\pi R)^p$. Introducing the Gauss-Bonnet correction in the lower-dimensional theory, with dimensionful coupling $\alpha$, one finds \cite{Clunan:2004tb}
\begin{equation}\label{eq:GB_corr_entropy}
    S^{(1)}_d = c_d \left(\frac{M}{M_\text{Pl,d}}\right)^{\frac{d-2}{d-3}} \left(1 + \frac{2(d-2)(d-3) \alpha}{R_\text{BH}^2}\right) ,
\end{equation}
where $M = M_\text{Pl,d}^{d-2}R_\text{BH}^{d-3}$ defines the horizon size in $d$ dimensions. At the transition, insofar as $R$ is large, so is $R_\text{BH}$. Indeed, to leading order $R_\text{BH} \sim R = m^{-1}$ up to a prefactor \cite{Bedroya:2024uva}. The subleading correction from eq. \ref{eq:GB_corr_entropy} is
\begin{align}
    R_\text{BH} & \sim 2\pi \left(\frac{c_D}{c_d}\right)^{\frac{D-3}{p}} R \left(1 - 2(d-2)(d-3)\frac{\alpha}{R_\text{BH}^2}\right) \\
    & \sim 2\pi \left(\frac{c_D}{c_d}\right)^{\frac{D-3}{p}} R \left(1 - \frac{(d-2)(d-3)}{2\pi^2} \left(\frac{c_d}{c_D}\right)^{\frac{D-3}{p}} \frac{\alpha}{R^2}\right) .
\end{align}
Therefore, the black-hole scale $\Lambda_\text{BH} \equiv R_\text{BH}^{-1}$ has a positive correction. All in all, rescaling $\alpha$ to absorb the numerical factors, and ignoring the overall prefactor which does not affect the parametric scale,
\begin{equation}\label{eq:bh_scale_corrected}
    \Lambda_\text{BH} \sim m \left(1 + \alpha \, m^2\right) .
\end{equation}
Let us observe that, in our setting, $\alpha$ is moduli-dependent. Indeed, its leading behavior is captured by the expression \cite{Maldacena:1997de}
\begin{equation}
\label{eq_Scorr}
    S_\text{corr} = \frac{1}{96 \pi} \int c_{2i}\, {\rm Im}\, z^i \,{\rm Tr} R \wedge * R \, ,
\end{equation}
so that $\alpha \sim y$ is linear in the (imaginary part of the) moduli. In computing the inner product of eq. \ref{thepattern} replacing $m$ by $\Lambda_\text{BH}$, we will first treat $\alpha$ as a constant to highlight the differences with respect to the above expression. 

\subsection{Constant Gauss-Bonnet coupling}

As in section \ref{sec:isotropic_torus}, we define $m = \frac{1}{[-i(T-\bar{T})]^{\gamma/2}}$. We shall keep $\gamma$ general, although for constant $\alpha$ it will drop out of the calculation once more. This is not true for the correct moduli-dependent $\alpha$, as will become apparent shortly. \\

Keeping in mind that this time the computation is only valid asymptotically, the logarithmic gradient $\nabla \log \Lambda_\text{BH}$ takes the form
\begin{align}\label{eq:log_Lambda_BH}
    \nabla \log \Lambda_\text{BH} & = \nabla \log m + \frac{2\alpha m}{1+\alpha m^2}\nabla m \\
    & = \left(1 + \frac{2\alpha m^2}{1+\alpha m^2}\right) \nabla \log m \sim \left(1+ 2\alpha m^2\right) \nabla \log m \, .
\end{align}
Hence, writing the asymptotic behavior of the inner product of eq. \ref{thepattern} according to
\begin{equation}
    \nabla \log m \cdot \nabla \log \Lambda_\text{sp} \sim \frac{1}{d-2} \left(1 + \epsilon \right) ,
\end{equation}
where the moduli-dependent relative correction $\epsilon$ was computed in the preceding section for isotropic limits of toroidal orbifolds, we find
\begin{equation}
    \nabla \log \Lambda_\text{BH} \cdot \nabla \log \Lambda_\text{sp} \sim \frac{1}{d-2} \left(1+2\alpha m^2\right)\left(1 + \epsilon \right) \sim \frac{1}{d-2}\left(1+\epsilon + 2\alpha m^2\right) .
\end{equation}
In other words, the relative correction brought by $\alpha$ behaves as $\frac{\alpha}{y^{\gamma}}$ and is positive. Therefore, at least for $\gamma=3$ where there are no additional directions of moduli space to account for, the leading correction is positive regardless of whether massless contributions are included. This would once more violate the proposed bound in eq. \ref{boundsbh} in this case; however, as we have shown in section \ref{sec:isotropic_torus}, the identification of eq. \ref{specscaletopstring} is not expected to hold for this limit. \\

For limits with $\gamma = 1$, we see from eq. \ref{eq:pattern_general_gamma} that the correction induced by the black-hole scale, which behaves as $\frac{1}{y^2}$, is subleading with respect to the (negative) one arising from the fixed tori. This is analogous to the K3 case in eq. \ref{eq:K3_corrections}. Therefore, we would be led once more to conclude that, at least for the toroidal orbifold studied in section \ref{sec:isotropic_torus}, the proposed bound in eq. \ref{boundsbh} does not hold once corrections to the leading asymptotics are included. However, since the correct $\alpha$ is not constant, we need to include its moduli-dependence in the calculation.

\subsection{Moduli-dependent Gauss-Bonnet coupling}

Including the moduli-dependence of $\alpha$, the logarithmic gradient of the black-hole scale in eq. \ref{eq:log_Lambda_BH} becomes
\begin{align}\label{eq:log_Lambda_BH_moduli}
    \nabla \log \Lambda_\text{BH} & = \nabla \log m + \frac{\nabla(\alpha m^2)}{1+\alpha m^2} \\
    & \sim \left(1+ 2\alpha m^2\right) \nabla \log m + m^2 \nabla \alpha \, .
\end{align}
Thus, in addition to the $\alpha m^2$ relative correction found in the preceding section, we also have the additional relative correction $2m^2 \nabla \alpha \cdot \nabla \log \Lambda_\text{sp}$. Since in four-dimensional Planck units $\alpha \sim F_1$ in the limit, this correction simplifies to
\begin{equation}
    - \, m^2 \, \frac{|\nabla F_1|^2}{F_1} \, .
\end{equation}
The Gauss-Bonnet coupling scales linearly in $y$ in the limit. This means that this negative quantity scales as $\frac{1}{y^2}$, still not enough to overcome the corrections from the extra moduli. Once again, this would naively lead us to conclude that the bound in eq. \ref{boundsbh} does not hold. However, for $\gamma=1$ the combination $\alpha m^2$ is not small at infinite distance, and our very starting point is not a reliable approximation. This is because in emergent string limits the tower mass gap coincides with the species scale (as defined in the introduction); hence, while the overall parametrics of black-hole thermodynamics is still reliable \cite{Basile:2023blg}, the detailed moduli dependence required to analyze gradients is not under control, at least for the techniques we employed. This is not surprising, since physical scales are only defined parametrically, whereas these slopes are sharp quantities. We thus learn that the cases in which the identification of the species scale according to eq. \ref{specscaletopstring} is expected to hold are precisely the ones in which the corrections to the black-hole scale cannot be reliably computed with the above approach! In other words, in the case of interest we cannot conclude whether eq. \ref{boundsbh} holds using the proper black-hole scale in eq. \ref{eq:bh_scale_corrected}.

\section{Conclusions}
\label{sec:conclusions}

In this paper we have investigated the CRV pattern of \cite{Castellano:2023stg, Castellano:2023jjt} to subleading order in infinite-distance limits. We focused on type II toroidal orbifolds, where the topological free energy can be computed exactly thanks to modular invariance. Moreover, we studied the black-hole scale introduced in \cite{Bedroya:2024uva} in this context, showing that subleading corrections to the pattern are sensitive to the difference between the black-hole scale and the mass gap of the dominant tower. Another subtlety that we identified is the role of massless species. As pointed out in \cite{vandeHeisteeg:2023ubh, vandeHeisteeg:2023dlw}, the standard Wilsonian philosophy would dictate that only massive states be integrated out when counting contributions to higher-derivative Wilson coefficients. Removing the corresponding terms, the slope of the species scale as a function of moduli is bounded by $\frac{1}{d-2}$ in all the examples studied in \cite{vandeHeisteeg:2023ubh, vandeHeisteeg:2023dlw}. \\

In this paper we explored the same subtlety in the context of the CRV pattern, showing that the asymptotic value for the inner product of gradients can be approached from below, depending on the direction taken in moduli space, when the contributions from massless states are removed. Although, as also pointed out in \cite{vandeHeisteeg:2023dlw}, the resulting expressions are no longer duality invariant (in this case, modular invariant), we find that this procedure can lead to negative corrections to the asymptotic values, but not in all cases depending on what direction in moduli space is chosen. Since the relation at stake involves derivatives along possibly several directions in moduli space, finding bounds along the lines of \cite{vandeHeisteeg:2023ubh, vandeHeisteeg:2023dlw} further supports the close relation between the tower scale and species scale entailed by the emergent string conjecture. \\

Replacing the tower mass gap with the black-hole scale of \cite{Bedroya:2024uva}, the correct moduli dependence in the subleading terms relative to the KK scale is crucial in order for the inner product to respect the bound proposed in \cite{Bedroya:2024uva}. In the cases of interest the deviation between the black-hole scale and the tower mass gap are not under control, which opens up the possibility to cure the violation of the bound which occurs for the tower mass gap. \\

All in all, at this stage a deeper investigation of these scales appears necessary to understand their role in this pattern, and the physical meaning behind their differences. More generally, it would be interesting to further ground this proposal with bottom-up motivations \cite{Castellano:2023stg}, as well as systematically explore the possible relations linking relevant physical scales and their derivatives. A natural direction to do so would be to look for relations between gradients of the species scale and vacuum energy, extending the analysis of \cite{Basile:2024lcz} within the framework of string perturbation theory. Another interesting avenue to pursue is the study of the CRV pattern in the absence of exact moduli spaces, using the tools developed in \cite{Lust:2019zwm, Stout:2021ubb, Velazquez:2022eco, DeBiasio:2022zuh, Li:2023gtt, Basile:2023rvm, Palti:2024voy, Mohseni:2024njl} for more general spaces of vacua or field configurations. Connections between these important quantities and their field dependence are likely to contain a great deal of physical insight, and achieving a more in-depth understanding of them would constitute an essential step toward uncovering the physical principles underlying string/M-theory and assessing string universality in quantum gravity.

\section*{Acknowledgements}

We would like to thank C. Montella, N. Cribiori and D. L\"{u}st for discussions and collaboration in the early stages of this project, and A. Castellano for useful correspondence and feedback on the manuscript. G.S. would like to thank V. Errasti Díez for comments on an earlier version of the draft. The work of I.B. is supported by the Origins Excellence Cluster and the German-Israel-Project (DIP) on Holography and the Swampland.

\appendix

\section{Dedekind and Eisenstein functions and asymptotics}
\label{dedekind}

In this brief appendix we collect some useful properties of Dedekind and Eisenstein functions.

\subsection{Dedekind eta function}

The Dedekind eta function is defined as 
\begin{equation}
    \eta(\tau) = q^{1/24} \Pi^{\infty}_{n=1}(1-q^n), \qquad q=e^{2 \pi i \tau} \, .
\end{equation}
It is useful to recall the relation 
\begin{equation}
    d \frac{d}{dq} \text{log} \eta = \frac{1}{24} E_2(\tau) \, ,
\end{equation}
used multiple times in the main text. Using $\tau = \frac{1}{2 \pi i} \text{log}q$ and $\frac{d}{d \tau} = 2 \pi i q \frac{d}{dq}$, one obtains the simpler expression
\begin{equation}
     \frac{d}{d \tau} \eta(\tau) = \frac{ \pi i }{12} \eta(\tau) E_2(\tau) \, .
\end{equation}
In the axiodilatonic setup discussed in section \ref{sec:axiodilatonicsection} we follow the convention of \cite{Grimm:2007tm}, where the nome $q$ is defined as $q^{-S}$. This produces some factors of $2\pi$ relative to the standard convention, ultimately not affecting any physical quantity we compute.

\subsection{Asymptotics of the Dedekind eta function}

In order to extract the first subleading corrections to the inner product of logarithmic gradients in eq. \ref{thepattern}, it is sufficient to look at the first terms of the $q$-expansion of the Dedekind function. These are given by
\begin{equation}
    \begin{aligned}
        \eta(T) &\sim q^{1/24}(1-q-q^2) \, , \\
        \eta(\bar{T}) &\sim \bar{q}^{1/24}(1-\bar{q}-\bar{q}^2) \, ,
    \end{aligned}
\end{equation}
where we identify $T = x + iy$ as the modulus of the two-torus. We take $y$ large at fixed $x$. Thus, we have 
\begin{equation}
    \begin{aligned}
        \eta(T) &\sim e^{\pi i x/12} e^{-\pi  y/12}(1-e^{2 \pi i x}e^{- 2 \pi y}-(e^{2 \pi i x}e^{- 2 \pi y})^2) \, ,\\
        \eta(\bar{T}) &\sim e^{-\pi i x/12} e^{-\pi  y/12}(1-e^{-2 \pi i x}e^{- 2 \pi y}-(e^{-2 \pi i x}e^{- 2 \pi y})^2) \, .
    \end{aligned}
\end{equation}
The asymptotic behavior of the absolute square $\eta(T) \eta(\bar{T})$, which appears in the topological free energy $F_1$ of the models we consider in the main text, reads
\begin{equation}
    \begin{aligned}
        \eta(T) \eta(\bar{T}) \sim e^{- \pi y/6} (1 - e^{-2 \pi y} (e^{2 \pi ix} + e^{-2 \pi ix}) + e^{-4 \pi y} (1 - e^{4 \pi ix} - e^{-4 \pi ix})) \, ,
    \end{aligned}
\end{equation}
so that finally
\begin{equation}
    \begin{aligned}
        \eta^2(T) \eta^2(\bar{T}) & \sim e^{- \pi y/3} (1 - e^{-2 \pi y} (e^{2 \pi ix} + e^{-2 \pi ix}) + e^{-4 \pi y} (1 - e^{4 \pi ix} - e^{-4 \pi ix}))^2 \\
        &\sim e^{- \pi y/3}(1-2e^{-2 \pi y}(e^{2 \pi ix} + e^{-2 \pi ix}) + 2 e^{-4 \pi y} (1 - e^{4 \pi ix} - e^{-4 \pi ix}) \\
        &+ e^{-4 \pi y} (e^{2 \pi ix} + e^{-2 \pi ix})^2) \, .
    \end{aligned}
\end{equation}

\subsection{Eisenstein series}

The Eisenstein series we employ in the main text are defined by 

\begin{equation}
    E_{2n} (q) = 1 - \frac{4n}{B_{2n}} \sum^{\infty}_{k=1} \frac{k^{2n-1} q^k}{1-q^k}
\end{equation}
via the Bernoulli numbers $B_m$. The function $E_2$ is famously not automorphic under modular transformations; its covariant but non-holomorphic version is
\begin{equation}
    \hat{E_2}(\tau, \bar{\tau}) = E_2(\tau) - \frac{3}{\pi \text{Im}\tau} = E_2(\tau) - \frac{6}{\pi (\tau - \bar{\tau})} \, .
\end{equation}

\subsection{Asymptotics of the Eisenstein series}

Once more, in order to extract the first subleading corrections to the physical quantities we compute in the main text, it is sufficient to include the next-to-leading asymptotics for small $q$. One finds 
\begin{equation}
    \begin{aligned}
        E_2(q) &\sim 1 - 24(q + 3q^2) \, , \\
        E_2(\bar{q}) &\sim 1 - 24(\bar{q} + 3\bar{q}^2) \, ,
    \end{aligned}
\end{equation}
where again $q=e^{2 \pi i T}$ and $T = x + iy$. 
Thus, we obtain 
\begin{equation}
    \begin{aligned}
        E_2(T) &\sim 1 - 24(e^{2 \pi ix - 2 \pi y} + 3e^{4 \pi ix - 4 \pi y} ) \, , \\
        E_2(\bar{T}) &\sim 1 - 24(e^{-2 \pi ix - 2 \pi y}  + 3e^{-4 \pi ix - 4 \pi y} ) \, . 
    \end{aligned}
\end{equation}

\printbibliography

\end{document}